\documentclass[manuscript]{acmart}

\AtBeginDocument{%
  }

\copyrightyear{2024}
\acmYear{2024}
\setcopyright{rightsretained}
\acmConference[MOBILEHCI Adjunct '24]{26th International Conference on Mobile Human-Computer Interaction}{September 30-October 3, 2024}{Melbourne, VIC, Australia}
\acmBooktitle{26th International Conference on Mobile Human-Computer Interaction (MOBILEHCI Adjunct '24), September 30-October 3, 2024, Melbourne, VIC, Australia}
\acmDOI{10.1145/3640471.3680236}
\acmISBN{979-8-4007-0506-9/24/09}

\usepackage{graphicx} 
\usepackage{pdfpages}
\usepackage{multirow, makecell}
\usepackage{booktabs}
\usepackage{hyperref}
\usepackage{multirow}
\usepackage{multicol}
\usepackage{float}
\usepackage{natbib}
\usepackage{csquotes}
\usepackage[figuresright]{rotating}
\usepackage{tablefootnote}
\usepackage{xcolor}
\usepackage{geometry}
\begin{document}

\title{Cultural influence on RE activities: An extended analysis of state of the art}

\author{Chowdhury Shahriar Muzammel}
\email{s3987367@student.rmit.edu.au}
\authornotemark[1]
\affiliation{%
  \institution{RMIT University}
  \city{Melbourne}
  \state{Victoria}
  \country{Australia}
}

\author{Maria Spichkova}
\email{maria.spichkova@rmit.edu.au}
\affiliation{%
  \institution{RMIT University}
  \city{Melbourne}
  \state{Victoria}
  \country{Australia}
}

\author{James Harland}
\email{james.harland@rmit.edu.au}
\affiliation{%
  \institution{RMIT University}
  \city{Melbourne}
  \state{Victoria}
  \country{Australia}
}


\renewcommand{\shortauthors}{C.S. Muzammel et al.}

\begin{abstract}
 Designing mobile software that aligns with cultural contexts is crucial for optimizing human-computer interaction. Considering cultural influences is essential not only for the actual set of functional/non-functional requirements, but also for the whole Requirement Engineering (RE) process. Without a clear understanding of cultural influences on RE activities, it's hardly possible to elaborate a correct and complete set of requirements. 
 This research explores the impact of national culture on RE-related activities based on recent studies. We conducted a Systematic Literature Review (SLR) of studies published in 2019-2023 and compared them to an older SLR covering 2000-2018. We identified 17 relevant studies, extracted 33 cultural influences impacting RE activities, and mapped them to the Hofstede model, widely used for cultural analysis in software development research. Our work highlights the critical role of national culture in RE activities, summarizes current research trends, and helps practitioners consider cultural influences for mobile app/software development.\\
~\\
 \emph{Preprint. Accepted to the 26th International Conference on Mobile Human-Computer Interaction (MOBILEHCI Adjunct '24), September 30-October 3, 2024, Melbourne, Australia. ACM Digital Library. Final version to be published by ACM Digital Library (In Press).}

\end{abstract}




\maketitle

\section{Introduction}

Developing interactive mobile applications is crucial in today's software engineering landscape. Having a correct and complete set of functional and non-functional requirements is essential for designing and developing mobile applications that effectively cater to the needs and cultural preferences of interacting users. However, to identify effectively the UI and functionality preferences, we need to adjust Requirement Engineering (RE) activities to take into account cultural influences not only on the requirements but also on RE activities. RE is commonly viewed as the initial phase of the software development process, wherein informal ideas are refined and translated into a formal specification~\cite{pohl1996requirements}. Pandey et al.~\cite{5656776} outlined the five key activities of Requirements Engineering as Elicitation, Analysis, Specification, Validation, and Management. Over recent years, it has been explored how the culture of stakeholders might impact the software development processes, especially the activities related to RE. The recognition of cultural influence on RE-related activities has notably expanded, underscoring its pivotal role within the broader spectrum of the software development field.

Developing software/apps for diverse cultures is challenging due to cultural differences in technology use. Addressing these influences during the RE process helps identify cultural preferences, and incorporating these preferences during RE enhances the effectiveness of software development. Alsswey and Al-Samarraie~\cite{alsswey2021role} used cultural preferences to design a mobile health app UI for Arab users, leading to high satisfaction in usability tests. Braun and Clarke~\cite{oh2011cultural} found that different cultures have varying preferences for mobile software, UI design, and applications. Thus, considering cultural factors during RE is very much essential. Different regions have unique cultural influences that can lead to challenges in RE and software development due to temporal, organizational, social, and geographical disparities~\cite{nadeem2019requirement, saleem2019understanding}.
Javed et al.~\cite{9794738} listed 74 factors and categorized them into eight groups, with `cultural differences' being the second one. Philemon et al.~\cite{yalamu2021cultural} conducted a case study on national culture of Papua New Guinean (PNG), identifying 11 cultural factors impacting RE, six of which align with Hofstede's dimensions~\cite{hofstede2011dimensionalizing} and five unique to PNG.

Alsanoosy et al. \cite{alsanoosy2020cultural} conducted a Systematic Literature Review (SLR) to analyze the potential influence of culture on RE-related activities. The researchers identified 16 cultural influences, which were mapped into six cultural dimensions as per Hofstede's theory~\cite{hofstede2010hofstede,hofstede2011dimensionalizing}. The Hofstede theory has been widely used for the analysis of issues relating to national culture\footnote{According to Google Scholar, Hofstede’s work has been cited more than 243,000 times, retrieved 15/04/2024.}. The SLR of Alsanoosy et al. covers the time period until 2018, but over the last five years, many new research studies have emerged; these may indicate new research issues and trends. Moreover, their systematic analysis might be useful. Driven by the need to understand  the evolution of research on cultural impact in RE practices, we formulated the following research question (RQ):
\begin{itemize}
    \item{\textbf{RQ:} What is the current status of the field and publication trends compared to an earlier SLR investigating how national culture impacts RE practices?}

\end{itemize}

\emph{Contributions:} 
Our objective is to understand current research trends and identify recent cultural influences on RE-related activities. To achieve this, we expanded upon the SLR by Alsanoosy et al.~\cite{alsanoosy2020cultural}, extending its scope to updated settings over a different timeframe. We analyzed publications emerging after the previous SLR's timeframe (2000-2018). From January 2019 to December 2023, we identified 17 relevant studies (in 19 papers). These studies confirmed 17 national cultural influences on RE-related activities identified in the previous SLR~\cite{alsanoosy2020cultural}. Additionally, we identified 16 new national cultural influences. In total, we identified 31 national cultural influences and 2 organizational influences, mapping them to relevant RE-related activities and the Hofstede model~\cite{hofstede2010hofstede}. Notably, publication frequency increased significantly, from an average of one publication per year (2000-2018) to nearly four per year (2019-2023). 

Our analysis provides insights into how national cultures shape RE-related activities in software development. As mobile applications are designed to serve users from various cultural backgrounds, software developers should consider these cultural influences not only during the UI design phase but also throughout the requirements engineering phase. This will aid in effective collaboration with diverse stakeholders and team members during requirements engineering, preventing miscommunication, enhancing the development process, reducing ambiguities, and ensuring accurate requirements gathering.

\section{Background: Hofstede Model}
\label{sec:background}

A survey~\cite{hoehle2015espoused} of mobile users from four major countries (the USA, Germany, China, and India) indicates that cultural values moderately impact social media software usability, and the researchers have used the Hofstede model~\cite{hofstede2010hofstede,hofstede2011dimensionalizing} to identify those impacts. Another study~\cite{oh2011cultural} analyzes top smartphone apps in 10 countries, mapping results to Hofstede dimensions to identify cultural differences in smartphone use. Another study~\cite{alsswey2021role} investigated how national cultural preferences influence the UI design of mobile phones for Arab users, using Hofstede dimensions to develop design guidelines for mobile applications. So, as part of the background analysis, it is essential to understand how the Hofstede model~\cite{hofstede2010hofstede,hofstede2011dimensionalizing} operates in the context of national culture.

Hofstede introduced one of the most widely accepted definitions and models of culture in cross-cultural studies~\cite{kirkman2006quarter,chang2015knowledge} and is being adopted in many studies~\cite{lim2014investigating, chaipunyathat2022communication, khan2022cultural, thanasankit2002understanding} regarding IT, SE, and RE. Hofstede suggests that in any group of people, there are many different ways of thinking, but these differences follow a pattern that helps us understand them~\cite{hofstede2010hofstede}. Culture influences almost everything we do and think. It's something we learn from being part of a social group, and it includes values, assumptions, and beliefs that affect how group members behave. This means that people from the same culture can understand each other well, but they might find it harder to communicate with people from different cultures. Hofstede conducted an extensive comparison~\cite{hofstede2010hofstede} of over 80 cultures and developed a model that provides cultural profiles to explore how culture influences our values in the workplace. According to the latest data available on Hofstede's official website, known as {\textquotedblleft}\href{https://www.hofstede-insights.com/country-comparison-tool} {Hofstede-Insights}{\textquotedblright}, cultural profiles are now available for 102 countries. 

Hofstede model, along with the cultural profiles, assists researchers in understanding the similarities and differences between national cultures~\cite{fischer2012cultural}. Each nation possesses unique cultural profiles characterized by specific Hofstede scores across various dimensions. Hofstede's cultural dimensions theory identifies six key aspects that help define and compare cultures: 

\begin{itemize}
\item 
\textbf{Power Distance Index (PDI):} Acceptance of unequal power distribution within a society;
\item %
\textbf{Individualism/Collectivism (IDV):} Emphasises personal goals versus group harmony;
\item %
\textbf{Masculinity/Femininity (MAS):} Focuses on competitiveness versus cooperation;
\item 
\textbf{Uncertainty Avoidance Index (UAI):} How comfortable a society is with uncertainty;
\item %
\textbf{Long-/Short-term Orientation (LTO):} Values tradition and persistence versus immediate gratification; and
\item 
\textbf{Indulgence/Impulses (IND):} Degree of freedom in satisfying basic human desires.
\end{itemize}

Each Hofstede cultural dimension is assessed using a numerical scale between 0 and 100, which allows for comparisons between different countries or cultures. These dimensions provide a framework to navigate cultural differences by recognizing and respecting diverse cultural values and behaviors within different countries with their respective scores. 
Each dimension has three ranges~\cite{alsanoosy2020framework}. For instance, a low Power Distance Index (PDI) score suggests an egalitarian society, prioritizing equal rights and opportunities for all. A high Individualism (IDV) score indicates an individualist society valuing personal freedom and success over group cohesion. Similarly, a medium Masculinity (MAS) score balances traditional masculine traits with feminine values like cooperation. Alsanoosy also proposed a framework~\cite{alsanoosy2020framework}, which provides anticipated cultural influences for a particular country with Hofstede scores.

\section{Methodology}
\label{sec:methodology}

\begin{figure}[ht!]
   \centering
   \caption{SLR review process and outcome}
   \includegraphics[width=\textwidth]{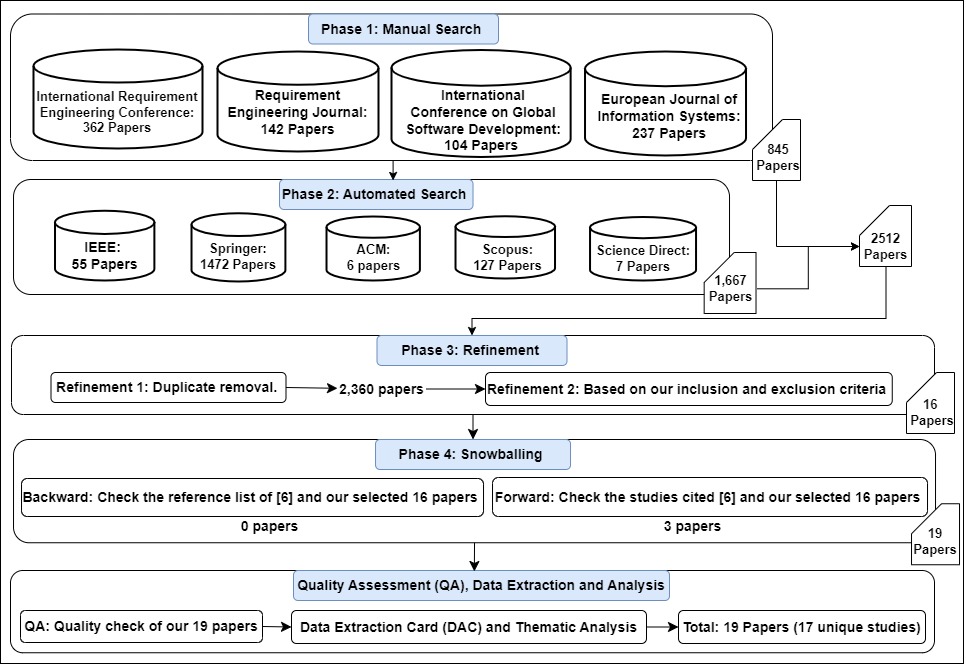}
   \label{fig:SLRSteps}
\end{figure}

In this section, we discuss the methodology of our study. To answer our RQ, we followed the methodology~\cite{alsanoosy2020cultural}, with the adjustments of the timeframes and other required settings. 
Figure \ref{fig:SLRSteps} illustrates the complete methodology of the SLR.

\textbf{Phase 1:} We applied the strategy of Zhang et al.~\cite{ZHANG2011625} by combining manual and automated search methods to get improved results. In the first phase, we manually checked different sources within a set time frame. This manual search aimed to create an unbiased search query using trusted studies rather than relying solely on our interpretations. The goal was to enhance the automated search's accuracy by finding and including more relevant studies. In the original SLR, the initial step was a manual search of four major publication venues for the period from January 2005 to December 2017. For RE, these journals and conferences are among the most popular venues for publication. Therefore, we conducted a manual search within the same conference proceedings and journals for the period spanning January 2018 to December 2023; see Figure~\ref{fig:SLRSteps}. 
At the end of phase 1, the manual search from 2018 to 2023 yielded 845 papers for review.

\textbf{Phase 2:} As the next step, we conducted an automatic search in IEEEXplore, ACM digital library, Springer, Science Direct, and Scopus. The construction of our search query involved analyzing the findings from the manual search and cross-referencing them with the search string used in the preceding SLR~\cite{alsanoosy2020cultural} on cultural influences. Upon evaluation, we found no need for alterations to the search string:

\emph{
(\enquote{requirements engineering} OR \enquote{requirements elicitation} OR \enquote{requirements gathering} OR \enquote{requirements identification} OR \enquote{requirements analysis} OR \enquote{requirements negotiation} OR \enquote{requirements specification} OR \enquote{requirements documentation} OR \enquote{requirements validation} OR {requirements management} OR \enquote{requirements change management}) AND (culture OR cultures OR cultural)
}.  
 
The only element we adjusted in this phase of SLR, was the time frame: 
while Alsanoosy et al. limited their search from January 2005 to December 2018, we conducted our search on publications January 2019 to December 2023. At the end of phase 2, from the automated search conducted from 2019 to 2023, we got 1,667 papers to review; see Figure~\ref{fig:SLRSteps}.

\textbf{Phase 3:} After manual and Automated search, we got 2,512 papers to review. To filter out our desired papers, we need to apply our intended specific inclusion criteria (IC) and exclusion criteria (EC). We will follow Kitchenham approach~\cite{kitchenham2004procedures} to filter our the desired papers. The criteria have been adapted from \cite{alsanoosy2020cultural}, with only time frames adjusted to be from January 2019 to December 2023. 
Thus, our ICs have been 
\begin{itemize}
    \item \textbf{(IC1)} Published in the English language,
    \item \textbf{(IC2)} Published between 
January 2019 and December 2023,
\item \textbf{(IC3)} Published in refereed journals, conferences, and workshops,
\item \textbf{(IC4)} Reporting national cultural influences influencing RE activities and/or reporting the industry practice of the RE process within a single culture.
\end{itemize}

Respectively, we applied the  following ECs:
\begin{itemize}
    \item 
    \textbf{(EC1)}  Published as an editorial, industrial experience, report, tutorial or keynote papers,
    \item 
    \textbf{(EC2)} Duplicated studies,
    \item 
    \textbf{(EC3)} Not focus explicitly on RE activities,
    \item 
\textbf{(EC4)} Not discuss explicitly  cultural influences influencing RE activities and/or how the RE process is practiced,
    \item 
    \textbf{(EC5)} Reporting organisational culture influences influencing RE activities.
\end{itemize}
According to EC2, we have removed 92 instances of duplicate papers, resulting in 2,360 unique papers to review. At the end of phase 3, after allying all other ICs and ECs, we got 16 papers.

\textbf{Phase 4:} We utilized reference searches, referred to as snowballing, to capture studies that could have been missed by both manual and automated searches due to limitations in certain electronic database search mechanisms. In the realm of SLRs, combining snowballing with database searches is recognized as an effective strategy for comprehensive coverage~\cite{wohlin2022successful}. We refined our SLR list by doing forward and backward snowballing iterations by the guidelines~\cite{wohlin2014guidelines}. 

\emph{Forward snowballing} entailed discovering additional studies by reviewing those that referenced our selected studies. 
\emph{Backward snowballing} involved identifying new studies by scrutinizing the reference lists of the chosen studies. 
We applied the Google Scholar data for both forward and backward snowballing. Our chosen studies were the 16 publications identified in Phase 3 of our SLR, and the publications identified in \cite{alsanoosy2020cultural}. As a result, we found three additional studies, all by forward snowballing. After this phase, we had a total of 19 papers (17 unique studies) for our SLR to analyse further.

\textbf{Quality Assessment:} 
In addition to the application of inclusion and exclusion criteria, 
we adopted the checklist by \cite{dybaa2008strength} for Quality Assessment (QA). We used a scoring technique to understand and assess the quality of the studies. All 19 papers qualified this assessment. We are identifying the final SLR list as: \\S1~\cite{yaseen2019validation}, S2~\cite{ALSANOOSY20192394}, S3~\cite{abd2019requirements}, S4d1~\cite{alsanoosy2019detailed}, S4d2~\cite{alsanoosy2019cultural}, S4~\cite{10.1145/3383219.3383266}, S5~\cite{alsanoosy2020framework}, S6~\cite{alsanoosy2020cultural}, S7~\cite{ALSANOOSY20203379}, S8~\cite{alsanoosy2020does}, S9~\cite{9270356}, S10~\cite{9206178}, S11~\cite{chaudhary2020res}, S12~\cite{tukur_umar_hassine_2021}, S13~\cite{alzayed2021requirements}, S14~\cite{9449943}, S15~\cite{uder2022}, S16~\cite{chaipunyathat2022communication}, and  S17~\cite{khan2022cultural}.

\textbf{Data Extraction and Analysis:} 
We applied Thematic Analysis (TA) \cite{guest2011applied} to the selected studies of our SLR to analyze data. For qualitative studies, TA is widely used for the data analysis process. We followed the TA analysis method of Braun and Clarke~\cite{braun2006using}. The first author reviewed all the studies, manually analyzed the relevant texts, labeled them, and generated the initial codes identifying cultural influences. If the labels directly complies with Hofstede's dimension \cite{hofstede2011dimensionalizing}, we matched the label accordingly. From those labels, we manually converted them into equivalent cultural influences as initial codes, which affect RE-related activities. Then, we unified the codes according to the cultural influences. We have identified 33 unique cultural influences as codes to map them into Hofstede's cultural model \cite{hofstede2011dimensionalizing}. At the same time, while doing the Thematic Analysis, we also created data extraction cards to extract information from our studies accurately following Kitchenham's process \cite{kitchenham2004procedures}.


\section{Results of the SLR}
\label{sec:results}

One of the aims of our work was to identify the current research trend. Since 2019, there has been a notable surge in publication frequency, rising from an average of one publication per year between 2000 and 2018 to nearly four publications per year between 2019 and 2023. Only over 2019 and 2020, there were 13 relevant published articles.

While analysing the relevant publications from 2019-2023, we identified 33 potential cultural influences, out of which 17 have been identified in the previous SLR~\cite{alsanoosy2020cultural} and 16 are newly identified by our study. To analyse how national culture impacts RE-related activities, we have mapped the identified cultural influences according to Hofstede's dimensions in Table~\ref{tab:IdentifiedCulturalInfluences}. 31 out of 33 influences are mapped to the Hofstede cultural dimensions, where two further influences have a different nature: they have been claimed by authors of the corresponding studies as culture-related, but they are less dependent on national culture while potentially related to organisational culture:
\begin{itemize}
    \item  \textit{(CI.32) Language and accent:} Different national languages and accents can also serve as organisational cultural barriers for conducting RE-related activities. Countries like Pakistan, South Korea, Saudi Arabia, and Australia, which host various multinational companies, encounter this issue as a hindrance to RE-related activities~\cite{yaseen2019validation, alsanoosy2019detailed, alsanoosy2019cultural, 9206178, chaudhary2020res, 9449943, uder2022, chaipunyathat2022communication} within an organisation.
 \item    
\textit{(CI.33) Loose employment of RE practices:} Practitioners from different nationalities sometimes take RE practices lightly, leading to impacts on RE-related activities within an organisation. Countries such as Saudi Arabia, Australia, Russia, the UK, Nigeria, and Gulf countries experience consequences related to this influence~\cite{alsanoosy2019cultural, tukur_umar_hassine_2021, alzayed2021requirements, 9449943}.

\end{itemize}

We also identified which cultural influences are affecting which RE-related activities from our selected studies in Table~\ref{tab:IdentifiedRE-relatedActivities}. Here, we identified five key stages of RE as: Elicitation (E), Analysis (A), Specification (S), Validation (V), and Management (M). We used the ``All'' column to indicate whether the specific culture influences all/overall RE-related activities or if the researchers did not explicitly mention any specific RE-related activities affected by it. To keep Table~\ref{tab:IdentifiedRE-relatedActivities} simple, we did not use the full names of cultural influences. Instead, we used the serial numbers of each cultural influence from Table~\ref{tab:IdentifiedCulturalInfluences}. We also used \textbf{bold} font to identify 16 newly recognized cultural influences in Table~\ref{tab:IdentifiedCulturalInfluences} and Table~\ref{tab:IdentifiedRE-relatedActivities}. Now, we are providing a brief description of each national cultural influence and discussing cultural influences identified within our work. We don't discuss here cultural influences that have been identified and discussed in the previous SLR~\cite{alsanoosy2020cultural} to focus on new influences and their impact.

\begin{table}[ht!]
    \centering
    \begin{scriptsize}
    \caption{Identified cultural influences from SLR studies}
    \label{tab:IdentifiedCulturalInfluences}
    \begin{tabular} {|c|l|l|}
\hline
\begin{tabular}[c]{@{}l@{}}Hofstede \\ Dimensions\end{tabular} & Cultural Influences  & Identified from studies \\ \hline

\multirow{11}{*}{PDI}    & (CI.1) Centralised decision-making & S2~\cite{ALSANOOSY20192394}, S4~\cite{10.1145/3383219.3383266},         
                        S5~\cite{alsanoosy2020framework}, S6~\cite{alsanoosy2020cultural}, S7~\cite{ALSANOOSY20203379}, S8~\cite{alsanoosy2020does}, S15~\cite{uder2022}                        \\ \cline{2-3} 
                        & (CI.2) Collaborative decision-making & S2~\cite{ALSANOOSY20192394}, S4~\cite{10.1145/3383219.3383266}, S5~\cite{alsanoosy2020framework}, S6~\cite{alsanoosy2020cultural}, S7~\cite{ALSANOOSY20203379}  \\ \cline{2-3} 
                        & (CI.3) Managers influence & S2~\cite{ALSANOOSY20192394}, S4d2~\cite{alsanoosy2019cultural}, S4~\cite{10.1145/3383219.3383266}, S5~\cite{alsanoosy2020framework}, S6~\cite{alsanoosy2020cultural},  S8~\cite{alsanoosy2020does}, S10~\cite{9206178}, S15~\cite{uder2022}, S17~\cite{khan2022cultural}  \\ \cline{2-3} 
                        &  (CI.4) Deference                    &  S2~\cite{ALSANOOSY20192394}, S4d1~\cite{alsanoosy2019detailed}, S4~\cite{10.1145/3383219.3383266}, S5~\cite{alsanoosy2020framework}, S6~\cite{alsanoosy2020cultural}, S8~\cite{alsanoosy2020does}, S15~\cite{uder2022},  S17~\cite{khan2022cultural}  \\ \cline{2-3} 
                        & (CI.5) Establish trust               &  S1~\cite{yaseen2019validation}, S2~\cite{ALSANOOSY20192394},  S4d1~\cite{alsanoosy2019detailed}, S4d2~\cite{alsanoosy2019cultural}, S4~\cite{10.1145/3383219.3383266}, S5~\cite{alsanoosy2020framework}, S6~\cite{alsanoosy2020cultural}, S10~\cite{9206178},  S17~\cite{khan2022cultural} \\ \cline{2-3} 
                        & \textbf{(CI.6)} Solving conflicts by favouritism or escalation  &  S2~\cite{ALSANOOSY20192394}, S4d1~\cite{alsanoosy2019detailed}, S4~\cite{10.1145/3383219.3383266}, S5~\cite{alsanoosy2020framework}, S7~\cite{ALSANOOSY20203379}, S8~\cite{alsanoosy2020does} \\ \cline{2-3} 
                        &  \textbf{(CI.7)} Solving conflicts by compromising  &  S2~\cite{ALSANOOSY20192394}, S4d2~\cite{alsanoosy2019cultural}, S4~\cite{10.1145/3383219.3383266}, S5~\cite{alsanoosy2020framework} \\ \cline{2-3} 
                        &  \textbf{(CI.8)} Avoiding conflicts  & S8~\cite{alsanoosy2020does} \\ \cline{2-3}
                        &  (CI.9) Hierarchical structure  & S6~\cite{alsanoosy2020cultural} \\ \cline{2-3}
                        &  \textbf{(CI.10)} Employees' attitude                       & S14~\cite{9449943} \\ \cline{2-3}
                        &  \textbf{(CI.11)} Punctuality & S7~\cite{ALSANOOSY20203379}  \\   \hline

\multirow{8}{*}{IDV}    &  (CI.12) Building relationships       & S4d1~\cite{alsanoosy2019detailed}, S4~\cite{10.1145/3383219.3383266},             
                        S5~\cite{alsanoosy2020framework}, S6~\cite{alsanoosy2020cultural}, S8~\cite{alsanoosy2020does}, S10~\cite{9206178}, S17~\cite{khan2022cultural} \\ \cline{2-3}
                        &  \textbf{(CI.13)} Taking ownership and responsibility  &  S4d2~\cite{alsanoosy2019cultural}, S4~\cite{10.1145/3383219.3383266}, S5~\cite{alsanoosy2020framework}, S6~\cite{alsanoosy2020cultural}    \\ \cline{2-3}
                        &  \textbf{(CI.14)} Openness and honesty         & S4d2~\cite{alsanoosy2019cultural}, S4~\cite{10.1145/3383219.3383266}, S5~\cite{alsanoosy2020framework}, S9~\cite{9270356}, S10~\cite{9206178},  S14~\cite{9449943}, S15~\cite{uder2022} \\ \cline{2-3}
                        & (CI.15) Safeguard workmates' jobs     &  S6~\cite{alsanoosy2020cultural} \\ \cline{2-3} 
                        & (CI.16) Communication context   & S1~\cite{yaseen2019validation}, S6~\cite{alsanoosy2020cultural},  
                        S10~\cite{9206178}, S11~\cite{chaudhary2020res}, S12~\cite{tukur_umar_hassine_2021}, S14~\cite{9449943}, S16~\cite{chaipunyathat2022communication} \\ \cline{2-3}
                        & (CI.17) Coordination and collaboration/Tearm work & S1~\cite{yaseen2019validation}, S6~\cite{alsanoosy2020cultural} \\ \cline{2-3}
                        & \textbf{(CI.18)} Hidden agenda                & S4d2~\cite{alsanoosy2019cultural}, S4~\cite{10.1145/3383219.3383266}, S5~\cite{alsanoosy2020framework}, S9~\cite{9270356} \\ \cline{2-3}
                        & \textbf{(CI.19)} Empathy with users & S4d1~\cite{alsanoosy2019detailed} \\ \hline

\multirow{3}{*}{MAS}    & (CI.20) Gender preference    & S6~\cite{alsanoosy2020cultural}, S11~\cite{chaudhary2020res},                
                        S17~\cite{khan2022cultural}       \\ \cline{2-3}
                        & (CI.21) Letting the strongest win    & S4d1~\cite{alsanoosy2019detailed}     \\ \cline{2-3}
                        & (CI.22) Gender segregation           & S3~\cite{abd2019requirements}, S4d1~\cite{alsanoosy2019detailed}, 
                        S6~\cite{alsanoosy2020cultural}, S7~\cite{ALSANOOSY20203379}  \\ \hline

\multirow{4}{*}{UAI}    & (CI.23) Recognition of uncertainty    & S4d1~\cite{alsanoosy2019detailed}, S4d2~\cite{alsanoosy2019cultural},                          
                        S4~\cite{10.1145/3383219.3383266}, S5~\cite{alsanoosy2020framework}, S6~\cite{alsanoosy2020cultural}, S8~\cite{alsanoosy2020does}, S12~\cite{tukur_umar_hassine_2021}  \\ \cline{2-3} 
                        & (CI.24) Subordinate avoid taking decision/risk & S4~\cite{10.1145/3383219.3383266}, S5~\cite{alsanoosy2020framework},  S7~\cite{ALSANOOSY20203379}, S17~\cite{khan2022cultural} \\ \cline{2-3} 
                        & \textbf{(CI.25)} Clients' resistance &  S4d2~\cite{alsanoosy2019cultural}, S5~\cite{alsanoosy2020framework}, S9~\cite{9270356}, S10~\cite{9206178}, S11~\cite{chaudhary2020res}, S13~\cite{alzayed2021requirements} \\ \cline{2-3} 
                        & \textbf{(CI.26)} Belief in expertise &  S4d1~\cite{alsanoosy2019detailed}               \\ \hline

\multirow{5}{*}{LTO}    & \textbf{(CI.27)} Aiming for quick results      & S4d2~\cite{alsanoosy2019cultural}, S4~\cite{10.1145/3383219.3383266},                          
                        S5~\cite{alsanoosy2020framework}, S7~\cite{ALSANOOSY20203379}, S14~\cite{9449943}  \\ \cline{2-3} 
                        & \textbf{(CI.28)} Solution-focused requirements   &  S4d2~\cite{alsanoosy2019cultural}  \\ \cline{2-3} 
                        & \textbf{(CI.29)} Relying on previous projects  &  S4~\cite{10.1145/3383219.3383266}, S5~\cite{alsanoosy2020framework}  \\ \cline{2-3}
                        & (CI.30) Face saving          &  S6~\cite{alsanoosy2020cultural} \\ \cline{2-3} 
                        & \textbf{(CI.31)} Dress code  & S4d1~\cite{alsanoosy2019detailed} \\ \hline

    \end{tabular}
    \end{scriptsize}
\end{table}

\begin{table}[ht!]
    \centering
    \begin{footnotesize}
    \caption{Identified cultural influences affecting RE-related activities}
    \label{tab:IdentifiedRE-relatedActivities}
    \begin{tabular}{|c|cccccc|}
\hline
\multicolumn{1}{|c|}{\multirow{2}{*}{\begin{tabular}[c]{@{}c@{}}Cultural\\Influences\end{tabular}}} &
  \multicolumn{6}{c|}{Impacted RE-related activities in identified studies} \\ \cline{2-7} 
\multicolumn{1}{|c|}{} &
  \multicolumn{1}{c|}{E} &
  \multicolumn{1}{c|}{A} &
  \multicolumn{1}{c|}{S} &
  \multicolumn{1}{c|}{V} &
  \multicolumn{1}{c|}{M} &
  All \\ \hline

  CI.1 &
  \multicolumn{1}{c|}{\cite{alsanoosy2019detailed, 10.1145/3383219.3383266}} &
  \multicolumn{1}{c|}{} &
  \multicolumn{1}{c|}{\cite{alsanoosy2019detailed, 10.1145/3383219.3383266, alsanoosy2020framework}} &
  \multicolumn{1}{c|}{\cite{10.1145/3383219.3383266, alsanoosy2020framework}} &  
  \multicolumn{1}{c|}{\cite{alsanoosy2019detailed, 10.1145/3383219.3383266, alsanoosy2020framework}} & \cite{ALSANOOSY20192394, alsanoosy2020cultural}  \\ \hline
 
  CI.2 &
  \multicolumn{1}{c|}{} &
  \multicolumn{1}{c|}{} &
  \multicolumn{1}{c|}{} &
  \multicolumn{1}{c|}{} &  
  \multicolumn{1}{c|}{} & \cite{ALSANOOSY20192394, 10.1145/3383219.3383266, alsanoosy2020framework, alsanoosy2020cultural} \\ \hline
 
  CI.3 &
  \multicolumn{1}{c|}{\cite{alsanoosy2019cultural, 10.1145/3383219.3383266, alsanoosy2020framework, alsanoosy2020cultural}} &
  \multicolumn{1}{c|}{} &
  \multicolumn{1}{c|}{\cite{alsanoosy2019cultural, 10.1145/3383219.3383266, alsanoosy2020framework}} &
  \multicolumn{1}{c|}{\cite{alsanoosy2019cultural, alsanoosy2020framework}} & 
  \multicolumn{1}{c|}{\cite{alsanoosy2019cultural, 10.1145/3383219.3383266, alsanoosy2020framework}} & \cite{ALSANOOSY20192394, alsanoosy2020cultural}
   \\ \hline
 
  CI.4 &
  \multicolumn{1}{c|}{\cite{alsanoosy2019detailed, 10.1145/3383219.3383266, alsanoosy2020framework, alsanoosy2020cultural}} &
  \multicolumn{1}{c|}{\cite{alsanoosy2019detailed}} &
  \multicolumn{1}{c|}{} &
  \multicolumn{1}{c|}{\cite{alsanoosy2019detailed, 10.1145/3383219.3383266, alsanoosy2020framework}} & 
  \multicolumn{1}{c|}{\cite{alsanoosy2019detailed}} & \cite{ALSANOOSY20192394, alsanoosy2020cultural}
   \\ \hline
 
  CI.5 &
  \multicolumn{1}{c|}{\cite{alsanoosy2019detailed, alsanoosy2019cultural, alsanoosy2020framework, alsanoosy2020cultural}} &
  \multicolumn{1}{c|}{\cite{alsanoosy2019detailed, alsanoosy2020cultural}} &
  \multicolumn{1}{c|}{} &
  \multicolumn{1}{c|}{\cite{alsanoosy2019detailed, alsanoosy2019cultural, alsanoosy2020framework}} & 
  \multicolumn{1}{c|}{\cite{alsanoosy2019detailed, alsanoosy2019cultural, alsanoosy2020framework}} & \cite{yaseen2019validation, ALSANOOSY20192394, 10.1145/3383219.3383266, alsanoosy2020cultural}
   \\ \hline 
 
  \textbf{CI.6} &
  \multicolumn{1}{c|}{\cite{10.1145/3383219.3383266}} &
  \multicolumn{1}{c|}{} &
  \multicolumn{1}{c|}{\cite{alsanoosy2020framework}} &
  \multicolumn{1}{c|}{\cite{10.1145/3383219.3383266, alsanoosy2020framework}} & 
  \multicolumn{1}{c|}{\cite{alsanoosy2020framework}} & \cite{ALSANOOSY20192394}
   \\ \hline
 
  \textbf{CI.7} &
  \multicolumn{1}{c|}{\cite{alsanoosy2019cultural, 10.1145/3383219.3383266, alsanoosy2020framework}} &
  \multicolumn{1}{c|}{} &
  \multicolumn{1}{c|}{\cite{alsanoosy2020framework}} &
  \multicolumn{1}{c|}{\cite{alsanoosy2019cultural, 10.1145/3383219.3383266, alsanoosy2020framework}} & 
  \multicolumn{1}{c|}{\cite{alsanoosy2019cultural, alsanoosy2020framework}} & \cite{ALSANOOSY20192394}
   \\ \hline
 
  \textbf{CI.8} &
  \multicolumn{1}{c|}{} &
  \multicolumn{1}{c|}{} &
  \multicolumn{1}{c|}{} &
  \multicolumn{1}{c|}{} &
  \multicolumn{1}{c|}{} & \cite{9449943}
   \\ \hline

  CI.9 &
  \multicolumn{1}{c|}{} &
  \multicolumn{1}{c|}{} &
  \multicolumn{1}{c|}{} &
  \multicolumn{1}{c|}{} & 
  \multicolumn{1}{c|}{} & \cite{alsanoosy2020cultural}
   \\ \hline 
 
  \textbf{CI.10} &
  \multicolumn{1}{c|}{} &
  \multicolumn{1}{c|}{} &
  \multicolumn{1}{c|}{} &
  \multicolumn{1}{c|}{} &
  \multicolumn{1}{c|}{} & \cite{9449943}
   \\ \hline
   
  \textbf{CI.11} &
  \multicolumn{1}{c|}{} &
  \multicolumn{1}{c|}{} &
  \multicolumn{1}{c|}{} &
  \multicolumn{1}{c|}{} &
  \multicolumn{1}{c|}{} & \cite{ALSANOOSY20203379}
   \\ \hline

  CI.12 &
  \multicolumn{1}{c|}{\cite{alsanoosy2019detailed, 10.1145/3383219.3383266, alsanoosy2020framework, alsanoosy2020cultural}} &
  \multicolumn{1}{c|}{} &
  \multicolumn{1}{c|}{\cite{alsanoosy2019detailed}} &
  \multicolumn{1}{c|}{\cite{10.1145/3383219.3383266, alsanoosy2020framework}} & 
  \multicolumn{1}{c|}{\cite{alsanoosy2019detailed, alsanoosy2020framework}} & \cite{alsanoosy2020cultural}\\ \hline
 
  \textbf{CI.13} &
  \multicolumn{1}{c|}{} &
  \multicolumn{1}{c|}{} &
  \multicolumn{1}{c|}{} &
  \multicolumn{1}{c|}{} & 
  \multicolumn{1}{c|}{} & \cite{alsanoosy2019cultural, 10.1145/3383219.3383266, alsanoosy2020framework}
   \\ \hline
 
  \textbf{CI.14} &
  \multicolumn{1}{c|}{\cite{alsanoosy2019cultural, 10.1145/3383219.3383266, alsanoosy2020framework}} &
  \multicolumn{1}{c|}{} &
  \multicolumn{1}{c|}{} &
  \multicolumn{1}{c|}{\cite{alsanoosy2019cultural, 10.1145/3383219.3383266, alsanoosy2020framework}} & 
  \multicolumn{1}{c|}{\cite{alsanoosy2019cultural, 10.1145/3383219.3383266, alsanoosy2020framework}} &
   \\ \hline
 
  CI.15 &
  \multicolumn{1}{c|}{\cite{alsanoosy2020cultural}} &
  \multicolumn{1}{c|}{} &
  \multicolumn{1}{c|}{} &
  \multicolumn{1}{c|}{} &
  \multicolumn{1}{c|}{} &
   \\ \hline
 
  CI.16 &
  \multicolumn{1}{c|}{\cite{alsanoosy2020cultural, chaudhary2020res, tukur_umar_hassine_2021}} &
  \multicolumn{1}{c|}{\cite{alsanoosy2020cultural, tukur_umar_hassine_2021}} &
  \multicolumn{1}{c|}{} &
  \multicolumn{1}{c|}{} &
  \multicolumn{1}{c|}{} & \cite{yaseen2019validation, 9449943}
   \\ \hline
 
  CI.17 &
  \multicolumn{1}{c|}{} &
  \multicolumn{1}{c|}{\cite{alsanoosy2020cultural}} &
  \multicolumn{1}{c|}{} &
  \multicolumn{1}{c|}{} &
  \multicolumn{1}{c|}{} & \cite{yaseen2019validation}
   \\ \hline
 
  \textbf{CI.18} &
  \multicolumn{1}{c|}{\cite{alsanoosy2019cultural, alsanoosy2020framework}} &
  \multicolumn{1}{c|}{\cite{alsanoosy2019cultural, alsanoosy2020framework}} &
  \multicolumn{1}{c|}{} &
  \multicolumn{1}{c|}{\cite{alsanoosy2019cultural}} & 
  \multicolumn{1}{c|}{\cite{alsanoosy2019cultural, alsanoosy2020framework}} & \cite{10.1145/3383219.3383266}
   \\ \hline

  \textbf{CI.19} &
  \multicolumn{1}{c|}{\cite{alsanoosy2019detailed}} &
  \multicolumn{1}{c|}{} &
  \multicolumn{1}{c|}{\cite{alsanoosy2019detailed}} &
  \multicolumn{1}{c|}{} & 
  \multicolumn{1}{c|}{\cite{alsanoosy2019detailed}} &
   \\ \hline 

  CI.20 &
  \multicolumn{1}{c|}{\cite{alsanoosy2020cultural, chaudhary2020res}} &
  \multicolumn{1}{c|}{} &
  \multicolumn{1}{c|}{} &
  \multicolumn{1}{c|}{} &
  \multicolumn{1}{c|}{} &
   \\ \hline
 
  CI.21 &
  \multicolumn{1}{c|}{\cite{alsanoosy2020cultural}} &
  \multicolumn{1}{c|}{\cite{alsanoosy2019detailed}} &
  \multicolumn{1}{c|}{} &
  \multicolumn{1}{c|}{\cite{alsanoosy2019detailed}} &
  \multicolumn{1}{c|}{} &
   \\ \hline

  CI.22 &
  \multicolumn{1}{c|}{\cite{alsanoosy2019detailed, alsanoosy2020cultural}} &
  \multicolumn{1}{c|}{} &
  \multicolumn{1}{c|}{\cite{alsanoosy2019detailed}} &
  \multicolumn{1}{c|}{} &
  \multicolumn{1}{c|}{} & \cite{abd2019requirements}
   \\ \hline

  CI.23 &
  \multicolumn{1}{c|}{\cite{alsanoosy2019cultural, alsanoosy2020cultural, tukur_umar_hassine_2021}} &
  \multicolumn{1}{c|}{\cite{alsanoosy2019cultural, alsanoosy2020cultural, tukur_umar_hassine_2021}} &
  \multicolumn{1}{c|}{\cite{alsanoosy2019cultural, alsanoosy2020cultural}} &
  \multicolumn{1}{c|}{} & 
  \multicolumn{1}{c|}{\cite{alsanoosy2019cultural}} & \cite{10.1145/3383219.3383266, alsanoosy2020framework, alsanoosy2020cultural}
   \\ \hline
 
  CI.24 &
  \multicolumn{1}{c|}{} &
  \multicolumn{1}{c|}{} &
  \multicolumn{1}{c|}{\cite{10.1145/3383219.3383266, alsanoosy2020framework}} &
  \multicolumn{1}{c|}{\cite{10.1145/3383219.3383266, alsanoosy2020framework}} &
  \multicolumn{1}{c|}{} & \cite{alsanoosy2020cultural}
   \\ \hline
 
  \textbf{CI.25} &
  \multicolumn{1}{c|}{\cite{alsanoosy2019cultural, chaudhary2020res, alzayed2021requirements}} &
  \multicolumn{1}{c|}{} &
  \multicolumn{1}{c|}{} &
  \multicolumn{1}{c|}{\cite{alsanoosy2019cultural}} & 
  \multicolumn{1}{c|}{\cite{alsanoosy2019cultural}} & \cite{alsanoosy2020framework}
   \\ \hline
 
  \textbf{CI.26} &
  \multicolumn{1}{c|}{\cite{alsanoosy2019detailed}} &
  \multicolumn{1}{c|}{\cite{alsanoosy2019detailed}} &
  \multicolumn{1}{c|}{} &
  \multicolumn{1}{c|}{} &
  \multicolumn{1}{c|}{} &
   \\ \hline

  \textbf{CI.27} &
  \multicolumn{1}{c|}{\cite{alsanoosy2019cultural}} &
  \multicolumn{1}{c|}{\cite{alsanoosy2019cultural}} &
  \multicolumn{1}{c|}{} &
  \multicolumn{1}{c|}{\cite{alsanoosy2019cultural}} & 
  \multicolumn{1}{c|}{\cite{alsanoosy2019cultural}} & \cite{10.1145/3383219.3383266, alsanoosy2020framework}
   \\ \hline
 
  \textbf{CI.28} &
  \multicolumn{1}{c|}{\cite{alsanoosy2019cultural}} &
  \multicolumn{1}{c|}{} &
  \multicolumn{1}{c|}{\cite{alsanoosy2019cultural}} &
  \multicolumn{1}{c|}{\cite{alsanoosy2019cultural}} & 
  \multicolumn{1}{c|}{\cite{alsanoosy2019cultural}} &
   \\ \hline
 
  \textbf{CI.29} &
  \multicolumn{1}{c|}{\cite{10.1145/3383219.3383266, alsanoosy2020framework}} &
  \multicolumn{1}{c|}{\cite{10.1145/3383219.3383266, alsanoosy2020framework}} &
  \multicolumn{1}{c|}{\cite{10.1145/3383219.3383266, alsanoosy2020framework}} &
  \multicolumn{1}{c|}{} &
  \multicolumn{1}{c|}{} &
   \\ \hline
 
  CI.30 &
  \multicolumn{1}{c|}{\cite{alsanoosy2020cultural}} &
  \multicolumn{1}{c|}{} &
  \multicolumn{1}{c|}{} &
  \multicolumn{1}{c|}{} &
  \multicolumn{1}{c|}{} & \cite{alsanoosy2020cultural}
   \\ \hline

  \textbf{CI.31} &
  \multicolumn{1}{c|}{\cite{alsanoosy2019detailed}} &
  \multicolumn{1}{c|}{} &
  \multicolumn{1}{c|}{\cite{alsanoosy2019detailed}} &
  \multicolumn{1}{c|}{} &
  \multicolumn{1}{c|}{} &
   \\ \hline

  CI.32 &
  \multicolumn{1}{c|}{\cite{alsanoosy2019cultural, alsanoosy2020cultural, chaudhary2020res}} &
  \multicolumn{1}{c|}{\cite{alsanoosy2020cultural}} &
  \multicolumn{1}{c|}{\cite{alsanoosy2020cultural}} &
  \multicolumn{1}{c|}{} &
  \multicolumn{1}{c|}{} & \cite{yaseen2019validation, alsanoosy2019detailed, alsanoosy2020cultural, 9449943}
   \\ \hline
 
  \textbf{CI.33} &
  \multicolumn{1}{c|}{} &
  \multicolumn{1}{c|}{} &
  \multicolumn{1}{c|}{} &
  \multicolumn{1}{c|}{} & 
  \multicolumn{1}{c|}{} & \cite{alsanoosy2019cultural, tukur_umar_hassine_2021, 9449943}
   \\ \hline

    \end{tabular}
    \end{footnotesize}
\end{table}

\subsubsection{\textbf{Power distance index}}
PDI measures how much people in a society accept authority and unequal power distribution~\cite{hofstede2010hofstede}. It reflects how comfortable individuals, especially those with less power, are with differences in authority and decision-making within institutions and organizations. We identified 11 cultural influences that can be mapped according to PDI. 
 \textit{(CI.3)}, \textit{(CI.4)}, \textit{(CI.5)}, \textit{(CI.9)} have been covered in the previous SLR~\cite{alsanoosy2020cultural}. Alsanoosy et al.~\cite{alsanoosy2020cultural} used ``Decision-making approaches'' to identify both ``Centralised'' and ``Collaborative decision-making'' approaches. However, we split them into two as they are two different types of decision-making approaches. 

\textit{(CI.1) Centralised decision-making:} This cultural influence refers how organisation take decisions. PDI impacts the level of centralization in control and decision-making processes~\cite{hofstede2010hofstede}. Countries with high PDI, like Saudi Arabia, face the negative impact of taking centralised decisions by their authorities~\cite{alsanoosy2020framework}.
    
\textit{(CI.2) Collaborative decision-making:} This cultural influence refers that authorities provide equal consideration when making decisions. In countries with low PDI, like Australia, authorities make decisions collaboratively, which has a mixed impact on RE-related activities~\cite{alsanoosy2020framework}.

\textit{(CI.6) Solving conflicts by favouritism or escalation:} This cultural influence relates to the method by which conflicts are resolved, often involving the use of favoritism. In Saudi Arabian culture, this method of conflict resolution is regarded as negative because it adversely affects RE-related activities~\cite{alsanoosy2020framework}.
    
\textit{(CI.7) Solving conflicts by compromising:} This cultural influence relates to compromise as a means of conflict resolution. In Australian culture, it positively influences RE-related activities~\cite{alsanoosy2020framework}.
    
\textit{(CI.8) Avoiding conflicts:} This cultural influence refers to how individuals simply avoid conflict to avoid any issues. In Vietnamese culture~\cite{alsanoosy2020does}, people tend to avoid conflict when making decisions regarding RE-related activities, which can impact the overall project.

\textit{(CI.10) Employees' attitude:} A survey~\cite{9449943} among requirement engineers from China and Germany shows that employees' attitudes, influenced by national culture, hinder the effective use of RE standards. This aligns with Hofstede's PDI dimension, where attitudes impact adherence to workplace rules, affecting RE-related activities.

\textit{(CI.11) Punctuality:} This pertains to the level of responsibility individuals from different cultures exhibit towards meeting deadlines and adhering to schedules. Stakeholders from Saudi Arabia experience a negative impact when they are not punctual in the traditional requirements engineering process~\cite{ALSANOOSY20203379}.

\subsubsection{\textbf{Individualism/collectivism}}
IDV measures the extent to which individuals prioritize their personal interests over the interests of the group or, conversely, prioritize the interests of the group over their own. We have identified 8 cultural influences that can be mapped according to IDV. The influences \textit{(CI.12)}, \textit{(CI.15)}, \textit{(CI.16)}, \textit{(CI.17)} have been covered in the previous SLR~\cite{alsanoosy2020cultural}.

\textit{(CI.13) Taking ownership and responsibility:} According to Hofstede et al.~\cite{hofstede2010hofstede}, this cultural attribute refers to individuals taking personal responsibility and accountability for their decision-making in a culture. In Australian culture, this is considered to be a positive cultural influence impacting RE-related activities~\cite{alsanoosy2020framework}.  
    
\textit{(CI.14) Openness and honesty:} In individualistic cultures, there is an encouragement for transparent sharing of emotions within the workplace~\cite{hofstede2010hofstede}. In Australian culture~\cite{alsanoosy2020framework}, this has a positive impact on RE-related activities.

\textit{(CI.18) Hidden agenda:} This cultural influence involves prioritizing individual interests over those of the group through the use of a hidden agenda. Individualistic cultures prioritize the needs of the individual over those of the group~\cite{hofstede2010hofstede}. Australian culture finds this attribute negatively impacting RE-related activities~\cite{alsanoosy2020framework}. 

\textit{(CI.19) Empathy with users:} This highlights the importance of client satisfaction. For Saudi Arabian practitioners, it's crucial that every change made fulfills their clients' needs~\cite{alsanoosy2019detailed}.

\subsubsection{\textbf{Masculinity/femininity}}
MAS measures the extent to which a society emphasizes traditionally masculine or feminine traits. 
Three cultural influences, \textit{(CI.20)}, \textit{(CI.21)}, and \textit{(CI.22)}, have been identified in the previous SLR~\cite{alsanoosy2020cultural}, no further MAS influences have been identified in the recent studies.

\subsubsection{\textbf{Uncertainity avoidance index}:}
UAI refers to how people handle various unknown situations~\cite{hofstede2010hofstede}. We identified 4 cultural influences that can be mapped under UAI. 
The influences \textit{(CI.23)} and \textit{(CI.24)} have been covered in the previous SLR~\cite{alsanoosy2020cultural}, we focus here on two other influences identified in the recent studies:

\textit{(CI.25) Clients’ resistance:} This cultural attribute refers to clients not cooperating properly during RE-related activities. Pakistani and Australian requirement engineers face challenges during such activities due to this issue, which impacts RE-related activities~\cite{alsanoosy2019cultural, alsanoosy2020framework, 9270356, 9206178, chaudhary2020res, alzayed2021requirements}.

\textit{(CI.26) Belief in expertise:} This influence refers to an individual's belief in and confidence in their expertise to accept recommendations. Clients in Saudi Arabia accept suggestions in RE-related activities because they consider them important, acknowledging that without trusting in expertise, successful outcomes would not be achievable~\cite{alsanoosy2019detailed}.

\subsubsection{\textbf{Long-/short-term orientation}:}
LTO refers to how much a culture values tradition and social obligations~\cite{hofstede2010hofstede}. We identified 5 cultural influences that can be mapped in LTO. We skip discussing \textit{(CI.30)}, as this was identified and discussed already in~\cite{alsanoosy2020cultural}.  

\textit{(CI.27) Aiming for quick results:} Cultures with low LTO focus on swift results to stay competitive~\cite{hofstede2010hofstede}.Consequently, individuals from these cultures in RE-related activities prioritize quick outcomes. Studies confirm that this cultural influence affects RE-related activities for various nations~\cite{alsanoosy2019cultural, 10.1145/3383219.3383266, alsanoosy2020framework, ALSANOOSY20203379, 9449943}.
    
\textit{(CI.28) Solution-focused requirements:} This influence refers to collecting requirements that primarily meet the users' needs. In Australian culture, this influence negatively affects RE-related activities~\cite{alsanoosy2019cultural}.  
    
\textit{(CI.29) Relying on previous projects:} Nations with low LTO maintain specific links to their history while navigating the demands of the present and future~\cite{hofstede2010hofstede}. Most requirement engineers from Saudi Arabia typically experience a positive impact from this cultural influence on RE-related activities~\cite{10.1145/3383219.3383266, alsanoosy2020framework}.

\textit{(CI.31) Dress code:}  The dress code for a specific gender is a traditional, social, and cultural practice. The implementation of a dress code in the workplace can impact RE-related activities if the client is uncomfortable with specific attire during the activity. In the Saudi Arabian cultural context, the implementation of a dress code can influence RE-related activities~\cite{alsanoosy2019detailed}.

Our research indicates that even there is an upward trend in the amount of publications per year on this topic. We identified 33 cultural influences, including 31 national and 2 organizational influences. We mapped the national cultural influences into Hofstede's dimensions~\cite{hofstede2010hofstede} and also identified how cultural influences affect RE-related activities. From these 33 cultural influences, 17 have already been identified by Alsanoosy et al. from their last SLR~\cite{alsanoosy2020cultural}, and the remaining 16 are newly identified from our selected studies compared to the previous SLR. In this section, we addressed our RQ by analysing the current publication trends and identified national cultural influences affecting RE-related activities based on our primary studies. 
Mobile applications are typically developed to meet the diverse needs of humans from different nationalities, which makes it essential to consider cultural influences not only on UI level but also on the level of RE-related activities. Effective interaction with diverse stakeholders and team members while doing RE will make the development process more efficient, allow to avoid misunderstanding and helps to gather correct and complete set of requirements.

\section{Threats to Validity}
\label{sec:threats}
Various factors could have influenced our evaluation. During the Thematic Analysis (TA) of the chosen case studies, we needed to extract cultural influences by coding information from the papers' raw data. It was challenging to determine if these cultural influences were tied to national cultures and if the data came from software engineering practitioners. Another challenge was several case studies dealt with multiple cultures, making it hard to distinguish specific cultural impacts. The first author conducted the coding process, which could potentially introduce bias if there is no further process on it. 
To mitigate this bias, the other two authors rechecked the raw data, reviewed the coding, and suggested necessary corrections. Then, all authors collaboratively reworked the coding to confirm they are indeed related to national cultural influences on RE for real-life scenarios. 

We conducted both forward and backward snowballing from the papers identified in our SLR. However, replicating the SLR inspired by the SLR by~\cite{alsanoosy2020cultural} means that solely snowballing from our chosen papers could lead us to miss relevant ones. To reduce this risk, we expanded our search by performing forward snowballing from the studies they identified in their SLR and backward snowballing from the papers they referenced. This effort resulted in finding three additional papers.

After a combined manual and automated search, we obtained 2,512 papers. The first author applied ICs and ECs to create the primary SLR list, posing another validity threat. To address this risk, the second author reviewed the outcome list and proposed corrections. Subsequently, the third author thoroughly reexamined the corrected list with all the papers to ensure proper application of the ICs and ECs.


\section{Conclusions}

\label{sec:concl}

Mobile software is designed to enable humans to interact with their mobile devices more effectively. Software is made by humans and made for humans to make life easier. In this process, a lot of human interaction and communication are needed, especially for RE-related activities, which is a fundamental step of building software. Previous studies confirmed that culture can impact these much-needed interactions and communications, and as a result, culture can impact RE-related activities. Therefore, our aim was to analyse the recent publication trends and to answer the following research question: 
\emph{RQ: What is the current status of the field and publication trends compared to an earlier SLR investigating how national culture impacts RE practices?}

To address the RQ, we explored the studies that have been conducted regarding cultural influences on RE-related activities since the beginning of 2019 
till December 2023. For this purpose, we reviewed 2,512 papers in total, and as a result, we identified 19 relevant publications or 17 studies. We noticed an upward trend in publications focusing on how national cultures affect RE-related activities lately, compared to before. After analyzing our identified studies, we found 33 cultural influences, including 31 national and 2 organisational influences that might impact RE-related activities, of which 16 are new compared to the previous SLR. We have mapped them onto Hofstede's dimensions and identified their impact on RE-related activities while developing software. Mobile apps/software are crafted to cater to the varied requirements of users worldwide of different cultures. Thus, integrating cultural factors not just in UI design but also throughout RE-related activities is important. Understanding the impacts of these identified cultural influences is vital, and they should be considered when working and interacting with team members of diverse nationalities and throughout all RE-related activities to prevent miscommunication and to enhance the accuracy of the RE process, which should aid in smoothly executing the software development process. This study benefits researchers, mobile and other software or mobile app developers, RE practitioners, policymakers, and educators by providing insights into the impacts of national cultural considerations in requirements engineering.

\bibliographystyle{ACM-Reference-Format}

\end{document}